\documentclass[aps,prl,reprint,groupedaddress,superscriptaddress,floatfix,10pt]{revtex4-1}

\usepackage{graphicx}

\begin{document}

\title{Time-Resolved Imaging of Negative Differential Resistance on the Atomic Scale}

\author{Mohammad Rashidi}
\email[Corresponding author.\\]{rashidi@ualberta.net}
\affiliation{Department of Physics, University of Alberta, Edmonton, Alberta, T6G 2J1, Canada.}
\affiliation{National Institute for Nanotechnology, National Research Council of Canada, Edmonton, Alberta, T6G 2M9, Canada.}
\author{Marco Taucer}
\affiliation{Department of Physics, University of Alberta, Edmonton, Alberta, T6G 2J1, Canada.}
\affiliation{National Institute for Nanotechnology, National Research Council of Canada, Edmonton, Alberta, T6G 2M9, Canada.}
\author{Isil Ozfidan}
\affiliation{Department of Physics, University of Alberta, Edmonton, Alberta, T6G 2J1, Canada.}
\author{Erika Lloyd}
\affiliation{Department of Physics, University of Alberta, Edmonton, Alberta, T6G 2J1, Canada.}
\author{Mohammad Koleini}
\affiliation{Department of Physics, University of Alberta, Edmonton, Alberta, T6G 2J1, Canada.}
\affiliation{National Institute for Nanotechnology, National Research Council of Canada, Edmonton, Alberta, T6G 2M9, Canada.}
\author{Hatem Labidi}
\affiliation{Department of Physics, University of Alberta, Edmonton, Alberta, T6G 2J1, Canada.}
\affiliation{National Institute for Nanotechnology, National Research Council of Canada, Edmonton, Alberta, T6G 2M9, Canada.}
\author{Jason L. Pitters}
\affiliation{National Institute for Nanotechnology, National Research Council of Canada, Edmonton, Alberta, T6G 2M9, Canada.}
\author{Joseph Maciejko}
\affiliation{Department of Physics, University of Alberta, Edmonton, Alberta, T6G 2J1, Canada.}
\affiliation{Canadian Institute for Advanced Research, Toronto, Ontario M5G 1Z8, Canada.}
\author{Robert A. Wolkow}
\affiliation{Department of Physics, University of Alberta, Edmonton, Alberta, T6G 2J1, Canada.}
\affiliation{National Institute for Nanotechnology, National Research Council of Canada, Edmonton, Alberta, T6G 2M9, Canada.}

\begin{abstract}
Negative differential resistance remains an attractive  but elusive functionality, so far only finding niche applications. Atom scale entities have shown promising properties, but viability of device fabrication requires fuller understanding of electron dynamics than has been possible to date. Using an all-electronic time-resolved scanning tunneling microscopy technique  and a Green's function transport model, we study an isolated dangling bond on a hydrogen terminated silicon surface. A robust negative differential resistance feature is identified as a many body phenomenon related to occupation dependent electron capture by a single atomic level. We measure all the time constants involved in this process and present atomically resolved, nanosecond timescale images to simultaneously capture the spatial and temporal variation of the observed feature.  
\end{abstract}

\maketitle

Half a century ago, William Shockley, upon hearing of Esaki's new diode and the strange negative differential resistance (NDR) it displayed~\cite{Esaki1958}, declared the Esaki diode, or tunnel diode, would play a dominant role in the then rapidly emerging field of semiconductor electronics. However, challenges related to complementary metal-oxide-semiconductor (CMOS) fabrication compatibility have persistently blocked the broad application of tunnel diodes. Because of the continued attraction of reduced device count and improved performance that NDR offers~\cite{Seabaugh1998}, researchers have continued to this day to seek physical mechanisms that yield NDR~\cite{Galperin2008,Perrin2014,Di2012,Wu2012,Choi2015,Lin2015}. Indications of NDR at the nanoscale have been enticing~\cite{Bedrossian1989,Lyo1989} but resisted reduction to practical implementation. Molecular-scale candidate assemblies have likewise fallen short of practical requirements and in some cases it has been revealed that molecular degradation occurring during measurement results in a current voltage trace that transiently masquerades as NDR only to fail shortly thereafter~\cite{Pitters2006,Tao2006}. Because of the false starts, a robust and well understood mechanism for NDR is desirable.

Well characterized atom scale ensembles, probed with the tip of a scanning tunneling microscope (STM) have recently led to renewed hope. One particular embodiment involves a silicon dangling bond (DB) on an otherwise boron terminated silicon (111) surface~\cite{Berthe2008,Nguyen2010}. Crucially, the observation of NDR is reproducible and does not require a chance unknown contamination of the metal scanned probe as had been the case in an early report~\cite{Lyo1989}. While a practical scalable fabrication process has not been reported, it is at least conceivable that one could be found. But before that practical step is taken, two more basic challenges must be overcome to create a robust basis for NDR. One issue is that DBs on the boron silicon surface occur at random during sample preparation and cannot be deliberately fabricated. Practical implementation requires control over placement of individual DBs and density of multiple DBs. Secondly, while a logical description of factors playing a role has been offered, no predictive theory has yet been established. 

Here, we address both of these issues. We show that DBs on a hydrogen-terminated Si(100) surface (H:Si), which can be patterned with atomic precision~\cite{Haider2009,Schofield2013,Pitters2011a}, exhibit the desired NDR. In addition to scanned probe-based current-voltage spectroscopy as has been used before, we apply an all-electronic time-resolved STM technique~\cite{Rashidi2016,Nunes1993,Loth2010,Loth2012,Grosse2013a,Saunus2013,Yan2014,Baumann2015} to isolate single atomic state carrier capture events, including measurement of an absolute capture rate on the nanosecond scale. A variation of that technique yields nanosecond scale temporal resolution without loss of atom-scale spatial imaging to reveal the lateral variation of the NDR effect. A Green's function transport model gives an account of the various rates participating in the process at different bias and tip height regimes. This model convincingly reproduces the measured data and thereby establishes a new mechanism to explain the physical process underlying the atom scale NDR process.

For these experiments, tungsten tips are electrochemically etched, cleaned, and sharpened by field ion microscopy~\cite{Rezeq2006a}. 
The samples are cleaved from a 3--4 m$\Omega.$cm n-type arsenic doped Si(100) wafer (Virginia Semiconductor Inc.) and are flash annealed at 1050$^o$C several times to remove the oxide layer. Hydrogen termination is done at 330$^o$C for 30 seconds under exposure to atomic hydrogen gas at 1$\times$10$^{-6}$~Torr.  The  measurements were performed at 4.5~K in ultra high vacuum using Omicron low temperature STM.  The Omicron STM is equipped commercially with radio frequency (RF) wiring that has a 500~MHz bandwidth. An arbitrary function generator (Tektronix AFG3252C) was used to generate a cycle of voltage pulses that are fed to the tip. An RF switch (Mini-Circuits ZX80-DR230-S+) is used to change the tip between ground and the output of the arbitrary function generator.

\begin{figure}
\includegraphics{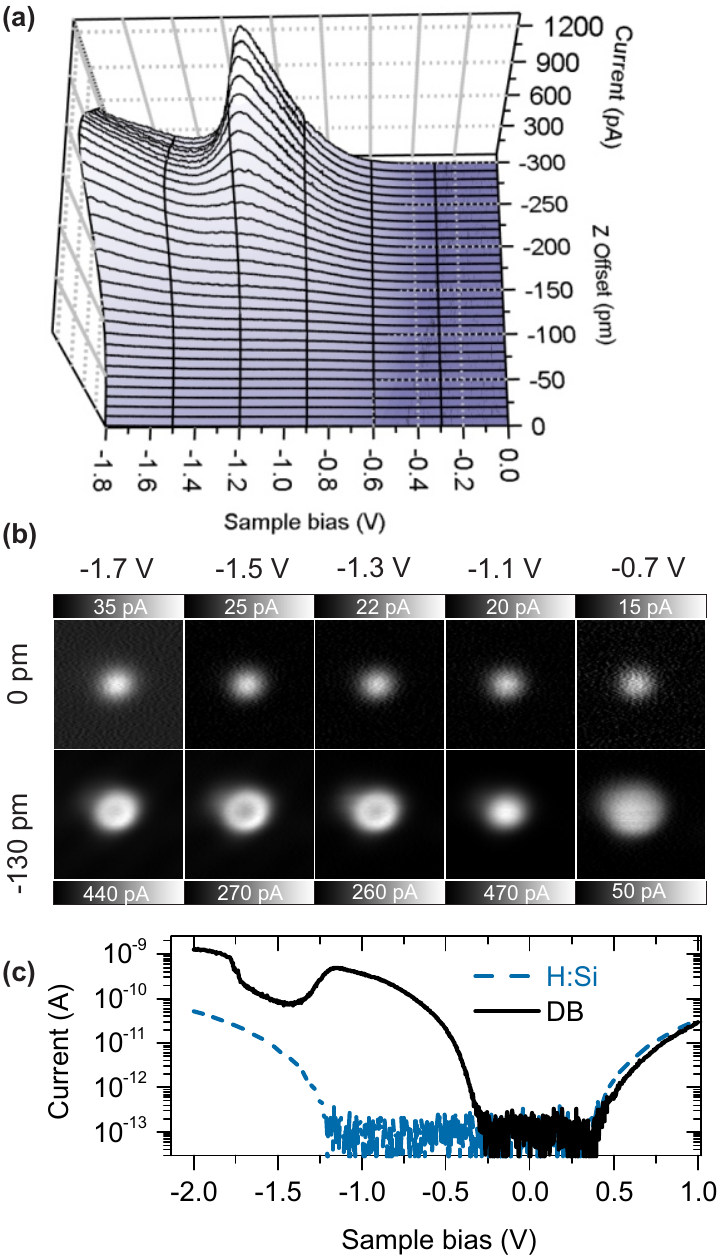}
\caption{\label{Fig1}  (a) Typical filled state \textit{I}(\textit{V},\textit{Z}) spectra of a single DB on highly doped hydrogen terminated silicon(100) recorded at  4.5 K with an STM probe at the DB center and at different tip-height offsets. The initial tip height (zero tip offset) is set with the tunneling conditions of  --1.80~V and 30.0~pA on top of the DB. The tip is moved subsequently 310~pm closer to the DB in 10~pm increments.  (b) Constant height STM images of the DB in (a) at different tip-height offsets (indicated on the left side) and sample bias voltages (indicated on top of each column). The tunneling current range for each STM image is indicated inside the color-bars attached to their frame. The size of the images is 3$\times$3 nm$^2$. The slight asymmetry in the images is due to an asymmetric tip shape.
(c)  Semi-log plots of \textit{I}(\textit{V}) spectra on a DB (different DB than the one shown in (a) and (b)) and H:Si at the same tip-height. }
\end{figure}

A DB has a single localized level in the bulk band gap that can hold up to two electrons. A DB with no electrons will be in a positive charge state (DB$^+$), a single electron puts the DB in a neutral charge state (DB$^0$), and the addition of a second electron brings the DB to a negative charge state (DB$^-$). The charge states, DB$^+$, DB$^0$, and DB$^-$, can be detected through their band bending effect on the surrounding H:Si~\cite{Taucer2014}, however, in STM \textit{I}(\textit{V}) spectroscopy, they are not directly observed. Instead, step-like features correspond to charge transition levels, denoted (+/0) and (0/--)~\cite{DelaBroise2000}. These levels correspond to the energies that must be imparted to an electron for it to drive a transition between charge states. The charge transition energies are due to polarization, relaxation, and in the case of (0/--) the on-site interaction energy, or Hubbard U, associated with the addition of an electron.

The measurements are performed on highly arsenic-doped H:Si heated to 1050$^o$C for oxide desorption, where negligible depletion of dopants near the surface is reported~\cite{Pitters2012}. Figure~\ref{Fig1}a reveals that unlike the results on samples with substantial dopant depletion region at the surface~\cite{Rashidi2016,Labidi2015}, the current onset arises in the bulk band gap. This indicates that the DB is being supplied by the bulk conduction band.  For close tip-surface distances, the tunneling spectra exhibit a current drop (NDR) at approximately  --1.20~V. For bias voltages more negative than the NDR onset, \textit{I}(\textit{Z}) spectra exhibit a peak, while for bias voltages less negative than the NDR onset, the tunneling current is exponentially dependent on the tip-height (Fig.~1a). Corresponding constant height STM images (Fig.~\ref{Fig1}b) reveal that for closer tip-surface distances the DB exhibits a dark center at bias voltages more negative than the NDR onset, indicating the spatial dependence of the NDR feature. 

 Figure~\ref{Fig1}c compares  \textit{I}(\textit{V}) spectra  over a DB and H:Si at the same tip height. The tunneling current onset for a DB occurs inside the bulk band gap, and for H:Si it occurs approximately at the energy of the silicon valence band edge. This indicates that in the NDR regime the dominant electron pathway from the conduction band to the tip is through the DB state as opposed to direct tunneling from the silicon conduction band to the tip. This pathway is likely mediated via vibronic coupling~\cite{Berthe2006}.

\begin{figure}
\includegraphics{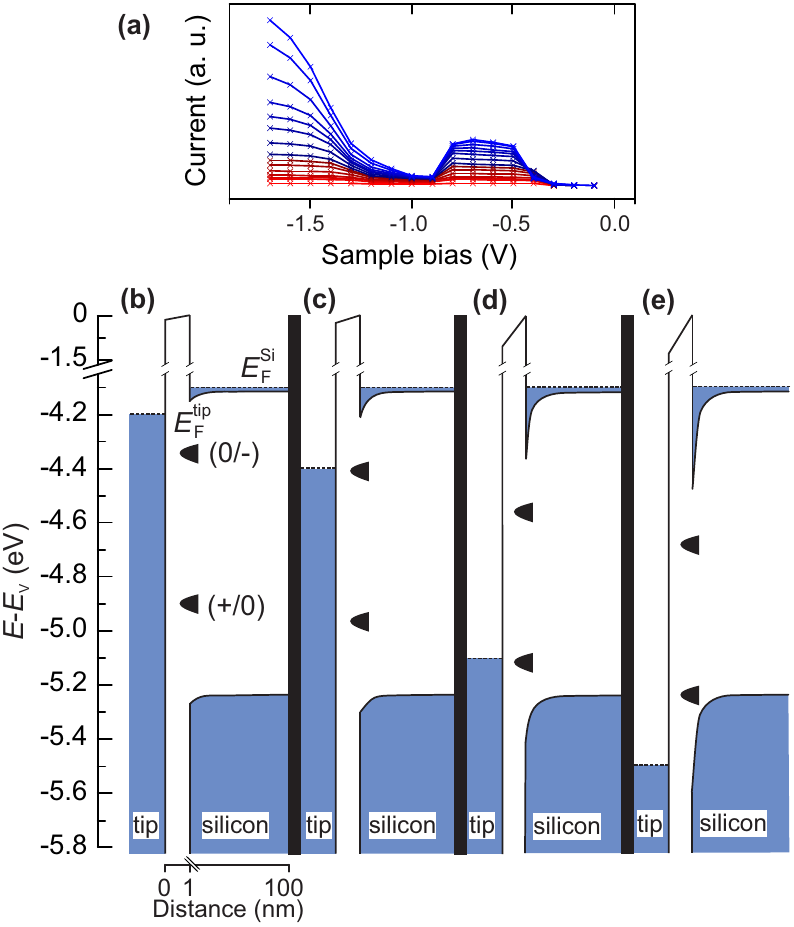}
\caption{\label{Fig2}  (a) Calculated \textit{I}(\textit{V}) spectra of a single silicon dangling bond at different tip-sample distances using NEGF. (b--e) Energy diagrams of the system of study for all the different energy regimes observed in the \textit{I}(\textit{V}) spectra. Fermi energy of the bulk silicon ($E_{\mathrm{F}}^{\mathrm{Si}}$) and the tip ($E_{\mathrm{F}}^{\mathrm{tip}}$) as well as the DB charge transition levels, (+/0) and (0/--), are displayed in the diagram. 
(b) The tip Fermi level is above both the DB charge transition levels. (c) The tip Fermi level is in resonance with  the (0/--)  level and above the (+/0) level.  (d) The tip Fermi level is in resonance with  the (+/0) level.  (e) The tip Fermi level is below both charge transition levels and due to tip induced band bending the (+/0) level becomes resonate with the bulk valence band edge.  }
\end{figure}

\begin{figure}
\includegraphics{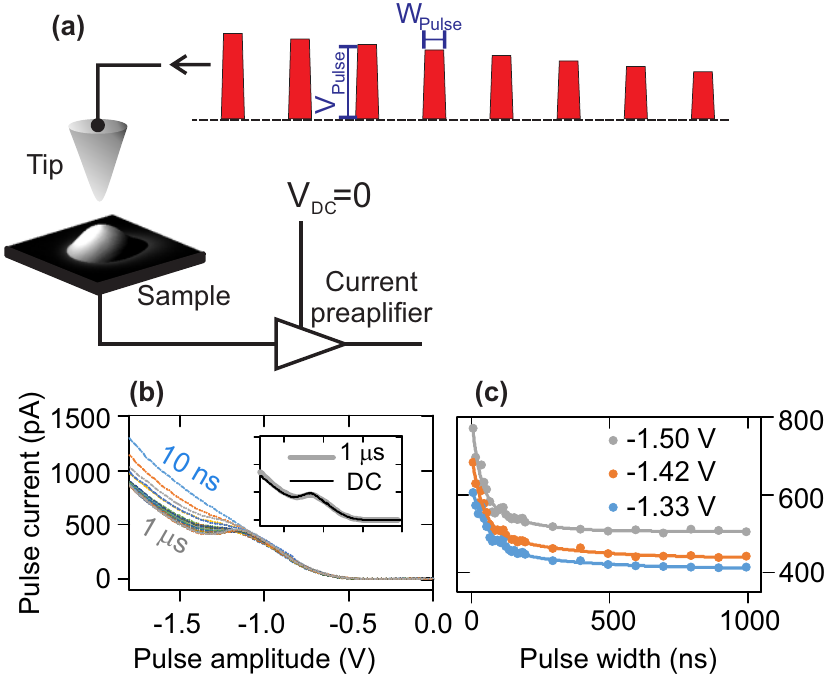}
\caption{\label{Fig3}  (a) Time-resolved \textit{I}(\textit{V}) spectroscopy measurement scheme. The DC bias is set to zero and the tunneling current is measured from a series of varying-amplitude nanosecond-regime pulses. (b) Time-resolved \textit{I}(\textit{V}) spectra recorded with tip-height offset of  --210 pm and different pulse widths (10~ns to 1~$\mu$s). The inset compares the DC \textit{I}(\textit{V}) and the time-resolved \textit{I}(\textit{V}) with 1~$\mu$s pulse width. The duty cycle for each curve is 0.05. The initial tip height (zero tip offset) is set with the tunneling conditions of  --1.80~V and 30.0~pA on top of the DB. (c) Pulse current as a function of pulse width for 3 different pulse amplitudes are shown as examples. Solid lines are exponential fits.}
\end{figure} 

We used a non-equilibrium Green's function (NEGF) method~\cite{Datta2005} to interpret our experimental results (Fig.~\ref{Fig2}a). In the NEGF formalism  coupling parameters between the DB and the tip ($\Gamma_\mathrm{tip}$) as well as  the DB and substrate ($\Gamma_\mathrm{Si}$) have been chosen to reproduced the experimental data~\cite{NEGF}. Details are available in the supplemental material.
These calculations agree with our experimental findings and allow us to understand the processes involved. 
The underlying mechanism for the observed characteristic \textit{I}(\textit{V}) curves is presented in the band diagrams (Fig.~\ref{Fig2}b--e),  separated into the four relevant energy regimes. 
Tip induced band bending is calculated for a 1D approximation using SEMITIP software~\cite{SEMITIP,INPUTPAR}. Since the tip work function is unknown, we assume a contact potential difference of zero, consistent with a previous study of the same surface~\cite{Pitters2012}.
The tunneling current is zero when the tip Fermi level is above the DB's charge transition levels (Fig.~\ref{Fig2}b). The current onset of \textit{I}(\textit{V}) spectra occurs when the tip Fermi level becomes resonant with the (0/--) level of the DB (Fig.~\ref{Fig2}c). The electrons start flowing from the conduction band through the (0/--) level of the DB and into the tip. As the sample bias becomes more negative, in agreement with the experiment, the NEGF calculations predict NDR. At the NDR onset, the tip becomes resonant with the (+/0) level (Fig.~\ref{Fig2}d), and the DB can occasionally be fully emptied by the tip. During the time that the DB is fully empty, the supply of electrons from the conduction band to the DB has to flow through the (+/0) level. Since this level is energetically much lower than the (0/--) level, the process is more inelastic and hence slow in comparison~\cite{Datta2005}. This results in NDR.  The NDR persists until the (+/0) level becomes resonant with the bulk valence band due to tip-induced band bending (Fig.~\ref{Fig2}e). As the sample bias becomes more negative, greater overlap between the DB and the bulk valence band allows for increased current flow. The separation between the higher and lower levels in a non-equilibrium setting is a many-body phenomenon. At a single particle level, ignoring the interactions, NDR would not have been present in the system. 

Due to the approximations made, the theory serves to explain the physical phenomenon leading to NDR qualitatively and not to reproduce exactly the measured \textit{I}(\textit{V}) spectra. While general features are well reproduced, the theory does not fully capture the experimental suppression of current in the very close tip regime. Specifically, while the experiment shows a reduction in the local current minimum at approximately  --1.4~V as the tip approaches the sample, in our theoretical calculations the minimum value of the current increases with the reduced tip-sample distance. This discrepancy may be a result of neglecting the dynamic interactions in self-energy which is expected to play a more important role as the distance between the tip and DB decreases.

We perform all-electronic time-resolved STM measurements~\cite{Rashidi2016,Nunes1993,Loth2010,Loth2012,Grosse2013a,Saunus2013,Yan2014,Baumann2015} to capture all the relevant rates in the NDR energy regime. A schematic of time-resolved \textit{I}(\textit{V}) spectroscopy measurement is shown in Fig.~\ref{Fig3}a. Time resolved measurements are distinct from their DC counterparts in that the tunneling current is induced by a series of short voltage pulses rather than a continuous DC bias. Pulse width, pulse amplitude, repetition rate and duty cycle can be tuned based on the conditions of the experiment. There exists an impedance mismatch between the 50~$\Omega$ of the outside circuit and the tunnel junction. In order to mitigate ringing at the tunnel junction, the pulses edges were kept to 2.5~ns. Furthermore, short pulses experience a distortion at the junction that creates an effective voltage at the junction that may be different than the applied voltage. To circumvent this, a series of fast \textit{I}(\textit{V}) curves for the pulse widths of interest were taken over H:Si, followed by a DC curve. The fast curves of the inert surface should reproduce the DC curve and an appropriate voltage shift can be extracted by comparison. This calibration reflects the difference in effective voltage and the applied voltage.  To calculate the current induced by the voltage pulses, we use $I_{\mathrm{preamp}}\times T/W=I_{\mathrm{eff}}$ formula. 
Where $T$ is the period of the pulse series, $W$ is the width of the pulse, $I_{\mathrm{preamp}}$ is  the measured current from the STM preamplifier and $I_{\mathrm{eff}}$ is the current induced by the voltage pulses.

Time-resolved \textit{I}(\textit{V}) measurements for a given tip height and different pulse widths are shown in Fig. 3b and compared with DC \textit{I}(\textit{V}). The NDR feature is extinguished for short pulses. The time-resolved \textit{I}(\textit{V}) spectra approach the DC \textit{I}(\textit{V}) for longer pulses (inset in Fig.~\ref{Fig3}b). The measured tunneling current for a given bias voltage exhibits an exponential decay with increasing pulse width (Fig.~\ref{Fig3}c).

\begin{figure}
\includegraphics[width=8.5cm]{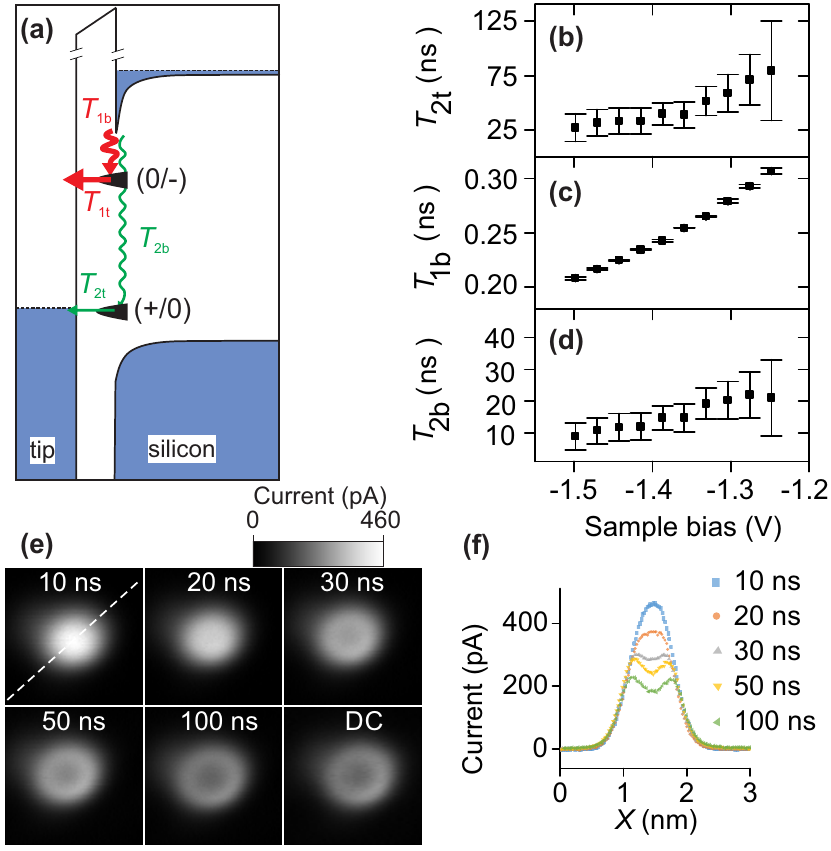}
\caption{\label{Fig4}  (a) Energy diagram of the system of study in the NDR regime displaying all the relevant time constants.  (b-d) The time constants displayed in (a) are extracted from the time-resolved \textit{I}(\textit{V}) spectra. (e) Time-resolved STM topographs. DC bias is kept at zero, pulse amplitude is  --1.5~V and the measurement duty cycle for each pulse width is 0.05. Image size is 2$\times$2 nm$^2$ and the tip-height offset is  --130~pm. DC image at  --1.5~V is added for comparison. The zero tip-offset is set with the tunneling conditions of --1.80~V and 30.0~pA on top of the DB. (f) Cross sectional profiles of the DB obtained from the time resolved image in (e). The white dashed line in (e) displays the direction where these profiles are extracted.}
\end{figure}

From the results in Fig.~\ref{Fig3}b and \ref{Fig3}c, we are able to capture all the time constants related to NDR (Fig.~\ref{Fig4}). The time constants of the exponential decays in Fig.~\ref{Fig3}c correspond to the time needed to fully empty the singly occupied DB by the tip ($T_{2\mathrm{t}}$). Figure~\ref{Fig4}b plots $T_{2\mathrm{t}}$ for different bias voltages in the NDR regime. During this time, the current flows only through the (0/--) level. Since the electron supply time constant from the bulk to the (0/--) level ($T_{1\mathrm{b}}$) is much longer than its emptying time by the tip ($T_{1\mathrm{t}}$), the measured current with the shortest pulse (10 ns) inversely corresponds to $T_{1\mathrm{b}}$. Therefore, we are able to extract $T_{1\mathrm{b}}$ from the spectrum of 10~ns width pulse (Fig.~\ref{Fig4}c). The time constant of electron supply from the bulk to the (+/0) level ($T_{2\mathrm{b}}$) can be calculated by (Supplementary Material):
	\[
	T_{2\mathrm{b}}(V)=\frac{\left[e-I_{\mathrm{DC}}(V)T_{1\mathrm{b}}(V)\right]T_{2\mathrm{t}}(V)}{T_{1\mathrm{b}}(V)I_{\mathrm{DC}}(V)},
\]
where $e$ is the elementary charge. Figure 4d shows $T_{2\mathrm{b}}$ for different bias voltages in the NDR regime. This is the direct measurement of electron capture time by a mid-gap state through an inelastic process.

The spatial evolution of a single DB at nanosecond time scales can be detected by performing time-resolved imaging (Fig.~\ref{Fig4}e). 
The resulting current maps are measured using short voltage pulses and resolve dynamics that occur on the time scale set by the pulse width. 
For the shortest pulses, the applied bias does not remain on long enough to fully empty the DB from the (+/0) level. Therefore, the electron transport occurs through the (0/--) level. Here, the DB appears as a simple protrusion (10~ns image in Fig.~\ref{Fig4}e and f). 
This simple protrusion corresponds to the hydrogen-like atomic orbital of the doubly occupied DB. 
As the pulse becomes longer, the conductivity at the center of the DB decreases (Fig.~\ref{Fig4}f) and the DB appearance approaches that of the DC measurements. This shows that the probability of emptying the DB from the (+/0) level is greater at its center, suggesting that the singly occupied DB is less extended than the doubly occupied DB, as expected.  This explains the observed ``ring'' shape of the DB in the NDR regime.

NDR continues to be a desirable but elusive functionality in electronic circuity. Here, we come to a detailed understanding of a structural and material configuration yielding robust NDR on silicon and which is therefore compatible with CMOS technology. As dangling bonds can be made en masse without need for scanned probe type tools, there appears to be the prospect for transitioning to a real technology that offers improved devices in the near future.
 

\begin{acknowledgments}
We would like to thank Martin Cloutier and Mark Salomons for their technical expertise, and Jacob Burgess and Sebastian Loth for stimulating discussions. We also thank NRC, NSERC, AITF, CRC, CIFAR and Compute Canada for support. 

\end{acknowledgments}


\begin{thebibliography}{35}%
\makeatletter
\providecommand \@ifxundefined [1]{%
 \@ifx{#1\undefined}
}%
\providecommand \@ifnum [1]{%
 \ifnum #1\expandafter \@firstoftwo
 \else \expandafter \@secondoftwo
 \fi
}%
\providecommand \@ifx [1]{%
 \ifx #1\expandafter \@firstoftwo
 \else \expandafter \@secondoftwo
 \fi
}%
\providecommand \natexlab [1]{#1}%
\providecommand \enquote  [1]{``#1''}%
\providecommand \bibnamefont  [1]{#1}%
\providecommand \bibfnamefont [1]{#1}%
\providecommand \citenamefont [1]{#1}%
\providecommand \href@noop [0]{\@secondoftwo}%
\providecommand \href [0]{\begingroup \@sanitize@url \@href}%
\providecommand \@href[1]{\@@startlink{#1}\@@href}%
\providecommand \@@href[1]{\endgroup#1\@@endlink}%
\providecommand \@sanitize@url [0]{\catcode `\\12\catcode `\$12\catcode
  `\&12\catcode `\#12\catcode `\^12\catcode `\_12\catcode `\%12\relax}%
\providecommand \@@startlink[1]{}%
\providecommand \@@endlink[0]{}%
\providecommand \url  [0]{\begingroup\@sanitize@url \@url }%
\providecommand \@url [1]{\endgroup\@href {#1}{\urlprefix }}%
\providecommand \urlprefix  [0]{URL }%
\providecommand \Eprint [0]{\href }%
\providecommand \doibase [0]{http://dx.doi.org/}%
\providecommand \selectlanguage [0]{\@gobble}%
\providecommand \bibinfo  [0]{\@secondoftwo}%
\providecommand \bibfield  [0]{\@secondoftwo}%
\providecommand \translation [1]{[#1]}%
\providecommand \BibitemOpen [0]{}%
\providecommand \bibitemStop [0]{}%
\providecommand \bibitemNoStop [0]{.\EOS\space}%
\providecommand \EOS [0]{\spacefactor3000\relax}%
\providecommand \BibitemShut  [1]{\csname bibitem#1\endcsname}%
\let\auto@bib@innerbib\@empty
\bibitem [{\citenamefont {Esaki}(1958)}]{Esaki1958}%
  \BibitemOpen
  \bibfield  {author} {\bibinfo {author} {\bibfnamefont {L.}~\bibnamefont
  {Esaki}},\ }\href {\doibase 10.1103/PhysRev.109.603} {\bibfield  {journal}
  {\bibinfo  {journal} {Physical Review}\ }\textbf {\bibinfo {volume} {109}},\
  \bibinfo {pages} {603} (\bibinfo {year} {1958})}\BibitemShut {NoStop}%
\bibitem [{\citenamefont {Seabaugh}\ \emph {et~al.}(1998)\citenamefont
  {Seabaugh}, \citenamefont {Deng}, \citenamefont {Blake}, \citenamefont
  {Brar}, \citenamefont {Broekaert}, \citenamefont {Lake}, \citenamefont
  {Morris},\ and\ \citenamefont {Frazier}}]{Seabaugh1998}%
  \BibitemOpen
  \bibfield  {author} {\bibinfo {author} {\bibfnamefont {A.}~\bibnamefont
  {Seabaugh}}, \bibinfo {author} {\bibfnamefont {X.}~\bibnamefont {Deng}},
  \bibinfo {author} {\bibfnamefont {T.}~\bibnamefont {Blake}}, \bibinfo
  {author} {\bibfnamefont {B.}~\bibnamefont {Brar}}, \bibinfo {author}
  {\bibfnamefont {T.}~\bibnamefont {Broekaert}}, \bibinfo {author}
  {\bibfnamefont {R.}~\bibnamefont {Lake}}, \bibinfo {author} {\bibfnamefont
  {F.}~\bibnamefont {Morris}}, \ and\ \bibinfo {author} {\bibfnamefont
  {G.}~\bibnamefont {Frazier}},\ }\href {\doibase 10.1109/IEDM.1998.746390}
  {\bibfield  {journal} {\bibinfo  {journal} {International Electron Devices
  Meeting 1998. Technical Digest}\ ,\ \bibinfo {pages} {429}} (\bibinfo {year}
  {1998})}\BibitemShut {NoStop}%
\bibitem [{\citenamefont {Galperin}\ \emph {et~al.}(2008)\citenamefont
  {Galperin}, \citenamefont {Ratner}, \citenamefont {Nitzan},\ and\
  \citenamefont {Troisi}}]{Galperin2008}%
  \BibitemOpen
  \bibfield  {author} {\bibinfo {author} {\bibfnamefont {M.}~\bibnamefont
  {Galperin}}, \bibinfo {author} {\bibfnamefont {M.~A.}\ \bibnamefont
  {Ratner}}, \bibinfo {author} {\bibfnamefont {A.}~\bibnamefont {Nitzan}}, \
  and\ \bibinfo {author} {\bibfnamefont {A.}~\bibnamefont {Troisi}},\ }\href
  {\doibase 10.1126/science.1146556} {\bibfield  {journal} {\bibinfo  {journal}
  {Science}\ }\textbf {\bibinfo {volume} {319}},\ \bibinfo {pages} {1056}
  (\bibinfo {year} {2008})}\BibitemShut {NoStop}%
\bibitem [{\citenamefont {Perrin}\ \emph {et~al.}(2014)\citenamefont {Perrin},
  \citenamefont {Frisenda}, \citenamefont {Koole}, \citenamefont {Seldenthuis},
  \citenamefont {Gil}, \citenamefont {Valkenier}, \citenamefont {Hummelen},
  \citenamefont {Renaud}, \citenamefont {Grozema}, \citenamefont {Thijssen},
  \citenamefont {Duli{\'{c}}},\ and\ \citenamefont {van~der
  Zant}}]{Perrin2014}%
  \BibitemOpen
  \bibfield  {author} {\bibinfo {author} {\bibfnamefont {M.~L.}\ \bibnamefont
  {Perrin}}, \bibinfo {author} {\bibfnamefont {R.}~\bibnamefont {Frisenda}},
  \bibinfo {author} {\bibfnamefont {M.}~\bibnamefont {Koole}}, \bibinfo
  {author} {\bibfnamefont {J.~S.}\ \bibnamefont {Seldenthuis}}, \bibinfo
  {author} {\bibfnamefont {J.~A.~C.}\ \bibnamefont {Gil}}, \bibinfo {author}
  {\bibfnamefont {H.}~\bibnamefont {Valkenier}}, \bibinfo {author}
  {\bibfnamefont {J.~C.}\ \bibnamefont {Hummelen}}, \bibinfo {author}
  {\bibfnamefont {N.}~\bibnamefont {Renaud}}, \bibinfo {author} {\bibfnamefont
  {F.~C.}\ \bibnamefont {Grozema}}, \bibinfo {author} {\bibfnamefont {J.~M.}\
  \bibnamefont {Thijssen}}, \bibinfo {author} {\bibfnamefont {D.}~\bibnamefont
  {Duli{\'{c}}}}, \ and\ \bibinfo {author} {\bibfnamefont {H.~S.~J.}\
  \bibnamefont {van~der Zant}},\ }\href {\doibase 10.1038/nnano.2014.177}
  {\bibfield  {journal} {\bibinfo  {journal} {Nature Nanotechnology}\ }\textbf
  {\bibinfo {volume} {9}},\ \bibinfo {pages} {830} (\bibinfo {year}
  {2014})}\BibitemShut {NoStop}%
\bibitem [{\citenamefont {Du}\ \emph {et~al.}(2012)\citenamefont {Du},
  \citenamefont {Pan}, \citenamefont {Wang}, \citenamefont {Wu}, \citenamefont
  {Feng}, \citenamefont {Pan},\ and\ \citenamefont {Wee}}]{Di2012}%
  \BibitemOpen
  \bibfield  {author} {\bibinfo {author} {\bibfnamefont {Y.}~\bibnamefont
  {Du}}, \bibinfo {author} {\bibfnamefont {H.}~\bibnamefont {Pan}}, \bibinfo
  {author} {\bibfnamefont {S.}~\bibnamefont {Wang}}, \bibinfo {author}
  {\bibfnamefont {T.}~\bibnamefont {Wu}}, \bibinfo {author} {\bibfnamefont
  {Y.~P.}\ \bibnamefont {Feng}}, \bibinfo {author} {\bibfnamefont
  {J.}~\bibnamefont {Pan}}, \ and\ \bibinfo {author} {\bibfnamefont {A.~T.~S.}\
  \bibnamefont {Wee}},\ }\href {\doibase 10.1021/nn204907t} {\bibfield
  {journal} {\bibinfo  {journal} {ACS Nano}\ }\textbf {\bibinfo {volume} {6}},\
  \bibinfo {pages} {2517} (\bibinfo {year} {2012})}\BibitemShut {NoStop}%
\bibitem [{\citenamefont {Wu}\ \emph {et~al.}(2012)\citenamefont {Wu},
  \citenamefont {Farmer}, \citenamefont {Zhu}, \citenamefont {Han},
  \citenamefont {Dimitrakopoulos}, \citenamefont {Bol}, \citenamefont
  {Avouris},\ and\ \citenamefont {Lin}}]{Wu2012}%
  \BibitemOpen
  \bibfield  {author} {\bibinfo {author} {\bibfnamefont {Y.}~\bibnamefont
  {Wu}}, \bibinfo {author} {\bibfnamefont {D.~B.}\ \bibnamefont {Farmer}},
  \bibinfo {author} {\bibfnamefont {W.}~\bibnamefont {Zhu}}, \bibinfo {author}
  {\bibfnamefont {S.~J.}\ \bibnamefont {Han}}, \bibinfo {author} {\bibfnamefont
  {C.~D.}\ \bibnamefont {Dimitrakopoulos}}, \bibinfo {author} {\bibfnamefont
  {A.~A.}\ \bibnamefont {Bol}}, \bibinfo {author} {\bibfnamefont
  {P.}~\bibnamefont {Avouris}}, \ and\ \bibinfo {author} {\bibfnamefont
  {Y.~M.}\ \bibnamefont {Lin}},\ }\href {\doibase 10.1021/nn205106z} {\bibfield
   {journal} {\bibinfo  {journal} {ACS Nano}\ }\textbf {\bibinfo {volume}
  {6}},\ \bibinfo {pages} {2610} (\bibinfo {year} {2012})}\BibitemShut
  {NoStop}%
\bibitem [{\citenamefont {Choi}\ \emph {et~al.}(2015)\citenamefont {Choi},
  \citenamefont {Lee}, \citenamefont {You}, \citenamefont {Lee},\ and\
  \citenamefont {Lee}}]{Choi2015}%
  \BibitemOpen
  \bibfield  {author} {\bibinfo {author} {\bibfnamefont {W.~S.}\ \bibnamefont
  {Choi}}, \bibinfo {author} {\bibfnamefont {S.~A.}\ \bibnamefont {Lee}},
  \bibinfo {author} {\bibfnamefont {J.~H.}\ \bibnamefont {You}}, \bibinfo
  {author} {\bibfnamefont {S.}~\bibnamefont {Lee}}, \ and\ \bibinfo {author}
  {\bibfnamefont {H.~N.}\ \bibnamefont {Lee}},\ }\href {\doibase
  10.1038/ncomms8424} {\bibfield  {journal} {\bibinfo  {journal} {Nature
  Communications}\ }\textbf {\bibinfo {volume} {6}},\ \bibinfo {pages} {7424}
  (\bibinfo {year} {2015})}\BibitemShut {NoStop}%
\bibitem [{\citenamefont {Lin}\ \emph {et~al.}(2015)\citenamefont {Lin},
  \citenamefont {Ghosh}, \citenamefont {Addou}, \citenamefont {Lu},
  \citenamefont {Eichfeld}, \citenamefont {Zhu}, \citenamefont {Li},
  \citenamefont {Peng}, \citenamefont {Kim}, \citenamefont {Li}, \citenamefont
  {Wallace}, \citenamefont {Datta},\ and\ \citenamefont {Robinson}}]{Lin2015}%
  \BibitemOpen
  \bibfield  {author} {\bibinfo {author} {\bibfnamefont {Y.-C.}\ \bibnamefont
  {Lin}}, \bibinfo {author} {\bibfnamefont {R.~K.}\ \bibnamefont {Ghosh}},
  \bibinfo {author} {\bibfnamefont {R.}~\bibnamefont {Addou}}, \bibinfo
  {author} {\bibfnamefont {N.}~\bibnamefont {Lu}}, \bibinfo {author}
  {\bibfnamefont {S.~M.}\ \bibnamefont {Eichfeld}}, \bibinfo {author}
  {\bibfnamefont {H.}~\bibnamefont {Zhu}}, \bibinfo {author} {\bibfnamefont
  {M.-Y.}\ \bibnamefont {Li}}, \bibinfo {author} {\bibfnamefont
  {X.}~\bibnamefont {Peng}}, \bibinfo {author} {\bibfnamefont {M.~J.}\
  \bibnamefont {Kim}}, \bibinfo {author} {\bibfnamefont {L.-J.}\ \bibnamefont
  {Li}}, \bibinfo {author} {\bibfnamefont {R.~M.}\ \bibnamefont {Wallace}},
  \bibinfo {author} {\bibfnamefont {S.}~\bibnamefont {Datta}}, \ and\ \bibinfo
  {author} {\bibfnamefont {J.~A.}\ \bibnamefont {Robinson}},\ }\href {\doibase
  10.1038/ncomms8311} {\bibfield  {journal} {\bibinfo  {journal} {Nature
  Communications}\ }\textbf {\bibinfo {volume} {6}},\ \bibinfo {pages} {7311}
  (\bibinfo {year} {2015})}\BibitemShut {NoStop}%
\bibitem [{\citenamefont {Bedrossian}\ \emph {et~al.}(1989)\citenamefont
  {Bedrossian}, \citenamefont {Chen}, \citenamefont {Mortensen},\ and\
  \citenamefont {Golovchenko}}]{Bedrossian1989}%
  \BibitemOpen
  \bibfield  {author} {\bibinfo {author} {\bibfnamefont {P.}~\bibnamefont
  {Bedrossian}}, \bibinfo {author} {\bibfnamefont {D.~M.}\ \bibnamefont
  {Chen}}, \bibinfo {author} {\bibfnamefont {K.}~\bibnamefont {Mortensen}}, \
  and\ \bibinfo {author} {\bibfnamefont {J.~A.}\ \bibnamefont {Golovchenko}},\
  }\href {\doibase 10.1038/342258a0} {\bibfield  {journal} {\bibinfo  {journal}
  {Nature}\ }\textbf {\bibinfo {volume} {342}},\ \bibinfo {pages} {258}
  (\bibinfo {year} {1989})}\BibitemShut {NoStop}%
\bibitem [{\citenamefont {Lyo}\ and\ \citenamefont {Avouris}(1989)}]{Lyo1989}%
  \BibitemOpen
  \bibfield  {author} {\bibinfo {author} {\bibfnamefont {I.~W.}\ \bibnamefont
  {Lyo}}\ and\ \bibinfo {author} {\bibfnamefont {P.}~\bibnamefont {Avouris}},\
  }\href {\doibase 10.1126/science.245.4924.1369} {\bibfield  {journal}
  {\bibinfo  {journal} {Science}\ }\textbf {\bibinfo {volume} {245}},\ \bibinfo
  {pages} {1369} (\bibinfo {year} {1989})}\BibitemShut {NoStop}%
\bibitem [{\citenamefont {Pitters}\ and\ \citenamefont
  {Wolkow}(2006)}]{Pitters2006}%
  \BibitemOpen
  \bibfield  {author} {\bibinfo {author} {\bibfnamefont {J.~L.}\ \bibnamefont
  {Pitters}}\ and\ \bibinfo {author} {\bibfnamefont {R.~A.}\ \bibnamefont
  {Wolkow}},\ }\href {\doibase 10.1021/nl0521569} {\bibfield  {journal}
  {\bibinfo  {journal} {Nano Lett.}\ }\textbf {\bibinfo {volume} {6}},\
  \bibinfo {pages} {390} (\bibinfo {year} {2006})}\BibitemShut {NoStop}%
\bibitem [{\citenamefont {Tao}(2006)}]{Tao2006}%
  \BibitemOpen
  \bibfield  {author} {\bibinfo {author} {\bibfnamefont {N.~J.}\ \bibnamefont
  {Tao}},\ }\href {\doibase 10.1038/nnano.2006.130} {\bibfield  {journal}
  {\bibinfo  {journal} {Nature Nanotechnology}\ }\textbf {\bibinfo {volume}
  {1}},\ \bibinfo {pages} {173} (\bibinfo {year} {2006})}\BibitemShut {NoStop}%
\bibitem [{\citenamefont {Berthe}\ \emph {et~al.}(2008)\citenamefont {Berthe},
  \citenamefont {Stiufiuc}, \citenamefont {Grandidier}, \citenamefont
  {Deresmes}, \citenamefont {Delerue},\ and\ \citenamefont
  {Sti{\'{e}}venard}}]{Berthe2008}%
  \BibitemOpen
  \bibfield  {author} {\bibinfo {author} {\bibfnamefont {M.}~\bibnamefont
  {Berthe}}, \bibinfo {author} {\bibfnamefont {R.}~\bibnamefont {Stiufiuc}},
  \bibinfo {author} {\bibfnamefont {B.}~\bibnamefont {Grandidier}}, \bibinfo
  {author} {\bibfnamefont {D.}~\bibnamefont {Deresmes}}, \bibinfo {author}
  {\bibfnamefont {C.}~\bibnamefont {Delerue}}, \ and\ \bibinfo {author}
  {\bibfnamefont {D.}~\bibnamefont {Sti{\'{e}}venard}},\ }\href {\doibase
  10.1126/science.1151186} {\bibfield  {journal} {\bibinfo  {journal}
  {Science}\ }\textbf {\bibinfo {volume} {319}},\ \bibinfo {pages} {436}
  (\bibinfo {year} {2008})}\BibitemShut {NoStop}%
\bibitem [{\citenamefont {Nguyen}\ \emph {et~al.}(2010)\citenamefont {Nguyen},
  \citenamefont {Mahieu}, \citenamefont {Berthe}, \citenamefont {Grandidier},
  \citenamefont {Delerue}, \citenamefont {Sti{\'{e}}venard},\ and\
  \citenamefont {Ebert}}]{Nguyen2010}%
  \BibitemOpen
  \bibfield  {author} {\bibinfo {author} {\bibfnamefont {T.~H.}\ \bibnamefont
  {Nguyen}}, \bibinfo {author} {\bibfnamefont {G.}~\bibnamefont {Mahieu}},
  \bibinfo {author} {\bibfnamefont {M.}~\bibnamefont {Berthe}}, \bibinfo
  {author} {\bibfnamefont {B.}~\bibnamefont {Grandidier}}, \bibinfo {author}
  {\bibfnamefont {C.}~\bibnamefont {Delerue}}, \bibinfo {author} {\bibfnamefont
  {D.}~\bibnamefont {Sti{\'{e}}venard}}, \ and\ \bibinfo {author}
  {\bibfnamefont {P.}~\bibnamefont {Ebert}},\ }\href {\doibase
  10.1103/PhysRevLett.105.226404} {\bibfield  {journal} {\bibinfo  {journal}
  {Physical Review Letters}\ }\textbf {\bibinfo {volume} {105}},\ \bibinfo
  {pages} {226404} (\bibinfo {year} {2010})}\BibitemShut {NoStop}%
\bibitem [{\citenamefont {Haider}\ \emph {et~al.}(2009)\citenamefont {Haider},
  \citenamefont {Pitters}, \citenamefont {Dilabio}, \citenamefont {Livadaru},
  \citenamefont {Mutus},\ and\ \citenamefont {Wolkow}}]{Haider2009}%
  \BibitemOpen
  \bibfield  {author} {\bibinfo {author} {\bibfnamefont {M.~B.}\ \bibnamefont
  {Haider}}, \bibinfo {author} {\bibfnamefont {J.~L.}\ \bibnamefont {Pitters}},
  \bibinfo {author} {\bibfnamefont {G.~A.}\ \bibnamefont {DiLabio}}, \bibinfo
  {author} {\bibfnamefont {L.}~\bibnamefont {Livadaru}}, \bibinfo {author}
  {\bibfnamefont {J.~Y.}\ \bibnamefont {Mutus}}, \ and\ \bibinfo {author}
  {\bibfnamefont {R.~A.}\ \bibnamefont {Wolkow}},\ }\href {\doibase
  10.1103/PhysRevLett.102.046805} {\bibfield  {journal} {\bibinfo  {journal}
  {Physical Review Letters}\ }\textbf {\bibinfo {volume} {102}},\ \bibinfo
  {pages} {046805} (\bibinfo {year} {2009})}\BibitemShut {NoStop}%
\bibitem [{\citenamefont {Schofield}\ \emph {et~al.}(2013)\citenamefont
  {Schofield}, \citenamefont {Studer}, \citenamefont {Hirjibehedin},
  \citenamefont {Curson}, \citenamefont {Aeppli},\ and\ \citenamefont
  {Bowler}}]{Schofield2013}%
  \BibitemOpen
  \bibfield  {author} {\bibinfo {author} {\bibfnamefont {S.~R.}\ \bibnamefont
  {Schofield}}, \bibinfo {author} {\bibfnamefont {P.}~\bibnamefont {Studer}},
  \bibinfo {author} {\bibfnamefont {C.~F.}\ \bibnamefont {Hirjibehedin}},
  \bibinfo {author} {\bibfnamefont {N.~J.}\ \bibnamefont {Curson}}, \bibinfo
  {author} {\bibfnamefont {G.}~\bibnamefont {Aeppli}}, \ and\ \bibinfo {author}
  {\bibfnamefont {D.~R.}\ \bibnamefont {Bowler}},\ }\href {\doibase
  10.1038/ncomms2679} {\bibfield  {journal} {\bibinfo  {journal} {Nature
  Communications}\ }\textbf {\bibinfo {volume} {4}},\ \bibinfo {pages} {1649}
  (\bibinfo {year} {2013})}\BibitemShut {NoStop}%
\bibitem [{\citenamefont {Pitters}\ \emph {et~al.}(2011)\citenamefont
  {Pitters}, \citenamefont {Livadaru}, \citenamefont {Haider},\ and\
  \citenamefont {Wolkow}}]{Pitters2011a}%
  \BibitemOpen
  \bibfield  {author} {\bibinfo {author} {\bibfnamefont {J.~L.}\ \bibnamefont
  {Pitters}}, \bibinfo {author} {\bibfnamefont {L.}~\bibnamefont {Livadaru}},
  \bibinfo {author} {\bibfnamefont {M.~B.}\ \bibnamefont {Haider}}, \ and\
  \bibinfo {author} {\bibfnamefont {R.~A.}\ \bibnamefont {Wolkow}},\ }\href
  {\doibase 10.1063/1.3514896} {\bibfield  {journal} {\bibinfo  {journal}
  {Journal of Chemical Physics}\ }\textbf {\bibinfo {volume} {134}},\ \bibinfo
  {pages} {064712} (\bibinfo {year} {2011})}\BibitemShut {NoStop}%
\bibitem [{\citenamefont {Rashidi}\ \emph {et~al.}(2016)\citenamefont
  {Rashidi}, \citenamefont {Burgess}, \citenamefont {Taucer}, \citenamefont
  {Achal}, \citenamefont {Pitters}, \citenamefont {Loth},\ and\ \citenamefont
  {Wolkow}}]{Rashidi2016}%
  \BibitemOpen
  \bibfield  {author} {\bibinfo {author} {\bibfnamefont {M.}~\bibnamefont
  {Rashidi}}, \bibinfo {author} {\bibfnamefont {J.}~\bibnamefont {Burgess}},
  \bibinfo {author} {\bibfnamefont {M.}~\bibnamefont {Taucer}}, \bibinfo
  {author} {\bibfnamefont {R.}~\bibnamefont {Achal}}, \bibinfo {author}
  {\bibfnamefont {J.~L.}\ \bibnamefont {Pitters}}, \bibinfo {author}
  {\bibfnamefont {S.}~\bibnamefont {Loth}}, \ and\ \bibinfo {author}
  {\bibfnamefont {R.~A.}\ \bibnamefont {Wolkow}},\ }\href@noop {} {\bibfield
  {journal} {\bibinfo  {journal} {Nature Communications}\ }\textbf {\bibinfo
  {volume} {7}},\ \bibinfo {pages} {13258} (\bibinfo {year}
  {2016})}\BibitemShut {NoStop}%
\bibitem [{\citenamefont {Nunes}\ and\ \citenamefont
  {Freeman}(1993)}]{Nunes1993}%
  \BibitemOpen
  \bibfield  {author} {\bibinfo {author} {\bibfnamefont {G.}~\bibnamefont
  {Nunes}}\ and\ \bibinfo {author} {\bibfnamefont {M.~R.}\ \bibnamefont
  {Freeman}},\ }\href {\doibase 10.1126/science.262.5136.1029} {\bibfield
  {journal} {\bibinfo  {journal} {Science}\ }\textbf {\bibinfo {volume}
  {262}},\ \bibinfo {pages} {1029} (\bibinfo {year} {1993})}\BibitemShut
  {NoStop}%
\bibitem [{\citenamefont {Loth}\ \emph {et~al.}(2010)\citenamefont {Loth},
  \citenamefont {Etzkorn}, \citenamefont {Lutz}, \citenamefont {Eigler},\ and\
  \citenamefont {Heinrich}}]{Loth2010}%
  \BibitemOpen
  \bibfield  {author} {\bibinfo {author} {\bibfnamefont {S.}~\bibnamefont
  {Loth}}, \bibinfo {author} {\bibfnamefont {M.}~\bibnamefont {Etzkorn}},
  \bibinfo {author} {\bibfnamefont {C.~P.}\ \bibnamefont {Lutz}}, \bibinfo
  {author} {\bibfnamefont {D.~M.}\ \bibnamefont {Eigler}}, \ and\ \bibinfo
  {author} {\bibfnamefont {A.~J.}\ \bibnamefont {Heinrich}},\ }\href {\doibase
  10.1126/science.1191688} {\bibfield  {journal} {\bibinfo  {journal}
  {Science}\ }\textbf {\bibinfo {volume} {329}},\ \bibinfo {pages} {1628}
  (\bibinfo {year} {2010})}\BibitemShut {NoStop}%
\bibitem [{\citenamefont {Loth}\ \emph {et~al.}(2012)\citenamefont {Loth},
  \citenamefont {Baumann}, \citenamefont {Lutz}, \citenamefont {Eigler},\ and\
  \citenamefont {Heinrich}}]{Loth2012}%
  \BibitemOpen
  \bibfield  {author} {\bibinfo {author} {\bibfnamefont {S.}~\bibnamefont
  {Loth}}, \bibinfo {author} {\bibfnamefont {S.}~\bibnamefont {Baumann}},
  \bibinfo {author} {\bibfnamefont {C.~P.}\ \bibnamefont {Lutz}}, \bibinfo
  {author} {\bibfnamefont {D.~M.}\ \bibnamefont {Eigler}}, \ and\ \bibinfo
  {author} {\bibfnamefont {A.~J.}\ \bibnamefont {Heinrich}},\ }\href {\doibase
  10.1126/science.1214131} {\bibfield  {journal} {\bibinfo  {journal}
  {Science}\ }\textbf {\bibinfo {volume} {335}},\ \bibinfo {pages} {196}
  (\bibinfo {year} {2012})}\BibitemShut {NoStop}%
\bibitem [{\citenamefont {Grosse}\ \emph {et~al.}(2013)\citenamefont {Grosse},
  \citenamefont {Etzkorn}, \citenamefont {Kuhnke}, \citenamefont {Loth},\ and\
  \citenamefont {Kern}}]{Grosse2013a}%
  \BibitemOpen
  \bibfield  {author} {\bibinfo {author} {\bibfnamefont {C.}~\bibnamefont
  {Grosse}}, \bibinfo {author} {\bibfnamefont {M.}~\bibnamefont {Etzkorn}},
  \bibinfo {author} {\bibfnamefont {K.}~\bibnamefont {Kuhnke}}, \bibinfo
  {author} {\bibfnamefont {S.}~\bibnamefont {Loth}}, \ and\ \bibinfo {author}
  {\bibfnamefont {K.}~\bibnamefont {Kern}},\ }\href {\doibase
  10.1063/1.4827556} {\bibfield  {journal} {\bibinfo  {journal} {Applied
  Physics Letters}\ }\textbf {\bibinfo {volume} {103}},\ \bibinfo {pages}
  {183108} (\bibinfo {year} {2013})}\BibitemShut {NoStop}%
\bibitem [{\citenamefont {Saunus}\ \emph {et~al.}(2013)\citenamefont {Saunus},
  \citenamefont {{Raphael Bindel}}, \citenamefont {Pratzer},\ and\
  \citenamefont {Morgenstern}}]{Saunus2013}%
  \BibitemOpen
  \bibfield  {author} {\bibinfo {author} {\bibfnamefont {C.}~\bibnamefont
  {Saunus}}, \bibinfo {author} {\bibfnamefont {J.}~\bibnamefont {{Raphael
  Bindel}}}, \bibinfo {author} {\bibfnamefont {M.}~\bibnamefont {Pratzer}}, \
  and\ \bibinfo {author} {\bibfnamefont {M.}~\bibnamefont {Morgenstern}},\
  }\href {\doibase 10.1063/1.4790180} {\bibfield  {journal} {\bibinfo
  {journal} {Applied Physics Letters}\ }\textbf {\bibinfo {volume} {102}},\
  \bibinfo {pages} {051601} (\bibinfo {year} {2013})}\BibitemShut {NoStop}%
\bibitem [{\citenamefont {Yan}\ \emph {et~al.}(2014)\citenamefont {Yan},
  \citenamefont {Choi}, \citenamefont {Burgess}, \citenamefont
  {Rolf-Pissarczyk},\ and\ \citenamefont {Loth}}]{Yan2014}%
  \BibitemOpen
  \bibfield  {author} {\bibinfo {author} {\bibfnamefont {S.}~\bibnamefont
  {Yan}}, \bibinfo {author} {\bibfnamefont {D.-J.}\ \bibnamefont {Choi}},
  \bibinfo {author} {\bibfnamefont {J.~A.~J.}\ \bibnamefont {Burgess}},
  \bibinfo {author} {\bibfnamefont {S.}~\bibnamefont {Rolf-Pissarczyk}}, \ and\
  \bibinfo {author} {\bibfnamefont {S.}~\bibnamefont {Loth}},\ }\href {\doibase
  10.1038/nnano.2014.281} {\bibfield  {journal} {\bibinfo  {journal} {Nature
  Nanotechnology}\ }\textbf {\bibinfo {volume} {10}},\ \bibinfo {pages} {40}
  (\bibinfo {year} {2014})}\BibitemShut {NoStop}%
\bibitem [{\citenamefont {Baumann}\ \emph {et~al.}(2015)\citenamefont
  {Baumann}, \citenamefont {Paul}, \citenamefont {Choi}, \citenamefont {Lutz},
  \citenamefont {Ardavan},\ and\ \citenamefont {Heinrich}}]{Baumann2015}%
  \BibitemOpen
  \bibfield  {author} {\bibinfo {author} {\bibfnamefont {S.}~\bibnamefont
  {Baumann}}, \bibinfo {author} {\bibfnamefont {W.}~\bibnamefont {Paul}},
  \bibinfo {author} {\bibfnamefont {T.}~\bibnamefont {Choi}}, \bibinfo {author}
  {\bibfnamefont {C.~P.}\ \bibnamefont {Lutz}}, \bibinfo {author}
  {\bibfnamefont {A.}~\bibnamefont {Ardavan}}, \ and\ \bibinfo {author}
  {\bibfnamefont {A.~J.}\ \bibnamefont {Heinrich}},\ }\href@noop {} {\bibfield
  {journal} {\bibinfo  {journal} {Science}\ }\textbf {\bibinfo {volume}
  {350}},\ \bibinfo {pages} {417} (\bibinfo {year} {2015})}\BibitemShut
  {NoStop}%
\bibitem [{\citenamefont {Rezeq}\ \emph {et~al.}(2006)\citenamefont {Rezeq},
  \citenamefont {Pitters},\ and\ \citenamefont {Wolkow}}]{Rezeq2006a}%
  \BibitemOpen
  \bibfield  {author} {\bibinfo {author} {\bibfnamefont {M.}~\bibnamefont
  {Rezeq}}, \bibinfo {author} {\bibfnamefont {J.~L.}\ \bibnamefont {Pitters}},
  \ and\ \bibinfo {author} {\bibfnamefont {R.}~\bibnamefont {Wolkow}},\ }\href
  {\doibase 10.1063/1.2198536} {\bibfield  {journal} {\bibinfo  {journal}
  {Journal of Chemical Physics}\ }\textbf {\bibinfo {volume} {124}},\ \bibinfo
  {pages} {204716} (\bibinfo {year} {2006})}\BibitemShut {NoStop}%
\bibitem [{\citenamefont {Taucer}\ \emph {et~al.}(2014)\citenamefont {Taucer},
  \citenamefont {Livadaru}, \citenamefont {Piva}, \citenamefont {Achal},
  \citenamefont {Labidi}, \citenamefont {Pitters},\ and\ \citenamefont
  {Wolkow}}]{Taucer2014}%
  \BibitemOpen
  \bibfield  {author} {\bibinfo {author} {\bibfnamefont {M.}~\bibnamefont
  {Taucer}}, \bibinfo {author} {\bibfnamefont {L.}~\bibnamefont {Livadaru}},
  \bibinfo {author} {\bibfnamefont {P.~G.}\ \bibnamefont {Piva}}, \bibinfo
  {author} {\bibfnamefont {R.}~\bibnamefont {Achal}}, \bibinfo {author}
  {\bibfnamefont {H.}~\bibnamefont {Labidi}}, \bibinfo {author} {\bibfnamefont
  {J.~L.}\ \bibnamefont {Pitters}}, \ and\ \bibinfo {author} {\bibfnamefont
  {R.~A.}\ \bibnamefont {Wolkow}},\ }\href@noop {} {\bibfield  {journal}
  {\bibinfo  {journal} {Physical Review Letters}\ }\textbf {\bibinfo {volume}
  {112}},\ \bibinfo {pages} {256801} (\bibinfo {year} {2014})}\BibitemShut
  {NoStop}%
\bibitem [{\citenamefont {de~la Bro{\"{i}}se}\ \emph
  {et~al.}(2000)\citenamefont {de~la Bro{\"{i}}se}, \citenamefont {Delerue},
  \citenamefont {Lannoo}, \citenamefont {Grandidier},\ and\ \citenamefont
  {Sti{\'{e}}venard}}]{DelaBroise2000}%
  \BibitemOpen
  \bibfield  {author} {\bibinfo {author} {\bibfnamefont {X.}~\bibnamefont
  {de~la Bro{\"{i}}se}}, \bibinfo {author} {\bibfnamefont {C.}~\bibnamefont
  {Delerue}}, \bibinfo {author} {\bibfnamefont {M.}~\bibnamefont {Lannoo}},
  \bibinfo {author} {\bibfnamefont {B.}~\bibnamefont {Grandidier}}, \ and\
  \bibinfo {author} {\bibfnamefont {D.}~\bibnamefont {Sti{\'{e}}venard}},\
  }\href {\doibase 10.1103/PhysRevB.61.2138} {\bibfield  {journal} {\bibinfo
  {journal} {Physical Review B}\ }\textbf {\bibinfo {volume} {61}},\ \bibinfo
  {pages} {2138} (\bibinfo {year} {2000})}\BibitemShut {NoStop}%
\bibitem [{\citenamefont {Pitters}\ \emph {et~al.}(2012)\citenamefont
  {Pitters}, \citenamefont {Piva},\ and\ \citenamefont {Wolkow}}]{Pitters2012}%
  \BibitemOpen
  \bibfield  {author} {\bibinfo {author} {\bibfnamefont {J.~L.}\ \bibnamefont
  {Pitters}}, \bibinfo {author} {\bibfnamefont {P.~G.}\ \bibnamefont {Piva}}, \
  and\ \bibinfo {author} {\bibfnamefont {R.~A.}\ \bibnamefont {Wolkow}},\
  }\href {\doibase 10.1116/1.3694010} {\bibfield  {journal} {\bibinfo
  {journal} {Journal of Vacuum Science {\&} Technology B}\ }\textbf {\bibinfo
  {volume} {30}},\ \bibinfo {pages} {021806} (\bibinfo {year}
  {2012})}\BibitemShut {NoStop}%
\bibitem [{\citenamefont {Labidi}\ \emph {et~al.}(2015)\citenamefont {Labidi},
  \citenamefont {Taucer}, \citenamefont {Rashidi}, \citenamefont {Koleini},
  \citenamefont {Livadaru}, \citenamefont {Pitters}, \citenamefont {Cloutier},
  \citenamefont {Salomons},\ and\ \citenamefont {Wolkow}}]{Labidi2015}%
  \BibitemOpen
  \bibfield  {author} {\bibinfo {author} {\bibfnamefont {H.}~\bibnamefont
  {Labidi}}, \bibinfo {author} {\bibfnamefont {M.}~\bibnamefont {Taucer}},
  \bibinfo {author} {\bibfnamefont {M.}~\bibnamefont {Rashidi}}, \bibinfo
  {author} {\bibfnamefont {M.}~\bibnamefont {Koleini}}, \bibinfo {author}
  {\bibfnamefont {L.}~\bibnamefont {Livadaru}}, \bibinfo {author}
  {\bibfnamefont {J.~L.}\ \bibnamefont {Pitters}}, \bibinfo {author}
  {\bibfnamefont {M.}~\bibnamefont {Cloutier}}, \bibinfo {author}
  {\bibfnamefont {M.}~\bibnamefont {Salomons}}, \ and\ \bibinfo {author}
  {\bibfnamefont {R.~A.}\ \bibnamefont {Wolkow}},\ }\href {\doibase
  10.1088/1367-2630/17/7/073023} {\bibfield  {journal} {\bibinfo  {journal}
  {New Journal of Physics}\ }\textbf {\bibinfo {volume} {17}},\ \bibinfo
  {pages} {073023} (\bibinfo {year} {2015})}\BibitemShut {NoStop}%
\bibitem [{\citenamefont {Berthe}\ \emph {et~al.}(2006)\citenamefont {Berthe},
  \citenamefont {Urbieta}, \citenamefont {Perdig{\~{a}}o}, \citenamefont
  {Grandidier}, \citenamefont {Deresmes}, \citenamefont {Delerue},
  \citenamefont {Sti{\'{e}}venard}, \citenamefont {Rurali}, \citenamefont
  {Lorente}, \citenamefont {Magaud},\ and\ \citenamefont
  {Ordej{\'{o}}n}}]{Berthe2006}%
  \BibitemOpen
  \bibfield  {author} {\bibinfo {author} {\bibfnamefont {M.}~\bibnamefont
  {Berthe}}, \bibinfo {author} {\bibfnamefont {A.}~\bibnamefont {Urbieta}},
  \bibinfo {author} {\bibfnamefont {L.}~\bibnamefont {Perdig{\~{a}}o}},
  \bibinfo {author} {\bibfnamefont {B.}~\bibnamefont {Grandidier}}, \bibinfo
  {author} {\bibfnamefont {D.}~\bibnamefont {Deresmes}}, \bibinfo {author}
  {\bibfnamefont {C.}~\bibnamefont {Delerue}}, \bibinfo {author} {\bibfnamefont
  {D.}~\bibnamefont {Sti{\'{e}}venard}}, \bibinfo {author} {\bibfnamefont
  {R.}~\bibnamefont {Rurali}}, \bibinfo {author} {\bibfnamefont
  {N.}~\bibnamefont {Lorente}}, \bibinfo {author} {\bibfnamefont
  {L.}~\bibnamefont {Magaud}}, \ and\ \bibinfo {author} {\bibfnamefont
  {P.}~\bibnamefont {Ordej{\'{o}}n}},\ }\href@noop {} {\bibfield  {journal}
  {\bibinfo  {journal} {Physical Review Letters}\ }\textbf {\bibinfo {volume}
  {97}},\ \bibinfo {pages} {206801} (\bibinfo {year} {2006})}\BibitemShut
  {NoStop}%
\bibitem [{\citenamefont {Datta}(2005)}]{Datta2005}%
  \BibitemOpen
  \bibfield  {author} {\bibinfo {author} {\bibfnamefont {S.}~\bibnamefont
  {Datta}},\ }\href@noop {} {\emph {\bibinfo {title} {{Quantum Transport: Atom
  to Transistor}}}}\ (\bibinfo  {publisher} {Cambridge University Press},\
  \bibinfo {year} {2005})\BibitemShut {NoStop}%
\bibitem [{NEG()}]{NEGF}%
  \BibitemOpen
  \href@noop {} {}\bibinfo {howpublished} {Input parameters:
  $\Gamma_\mathrm{tip}$=$3\times10^4\mathrm{exp}(-2.4d)$, where $d$ is the
  approximate tip-sample distance in nm.
  $\Gamma_\mathrm{Si}$=$0.082+0.005V_{\mathrm{bias}}$, where $V_\mathrm{bias}$
  is the sample bias voltage.}\BibitemShut {Stop}%
\bibitem [{SEM(2011)}]{SEMITIP}%
  \BibitemOpen
  \href@noop {} {}\bibinfo {howpublished}
  {SEMITIP:~\url{http://www.andrew.cmu.edu/user/feenstra/semitip_v6/}}
  (\bibinfo {year} {2011})\BibitemShut {NoStop}%
\bibitem [{INP()}]{INPUTPAR}%
  \BibitemOpen
  \href@noop {} {}\bibinfo {howpublished} {Input parameters: tip-sample
  separation= 1~nm, contact potential=0~eV, donor
  concentration=$2\times10^{19}$~cm$^{-3}$, dielectric constant=11.9, band
  gap=1.17~eV .}\BibitemShut {Stop}%
\end{thebibliography}
\end{document}